\documentclass{article}
\usepackage[numbers,sort&compress]{natbib}
\usepackage{color}
\usepackage{bbm,amsmath}
\usepackage{graphicx}
\usepackage{amssymb}
\usepackage{authblk}

\usepackage{lineno}
\usepackage{nameref,hyperref}
\usepackage{placeins}
\usepackage{todonotes}

\title{
Exploring and mapping the universe of evolutionary graphs
}
\author{
Marius M\"oller*, Laura Hindersin*, and Arne Traulsen}
\affil{
Department of Evolutionary Theory, Max Planck Institute for Evolutionary Biology, D-24306 Pl\"on, Germany.\\
{\small * equal contribution}
}
\begin{document}

\maketitle
\section*{Abstract}
Population structure can be modelled by evolutionary graphs, which can have a substantial, but very subtle influence on the fate of the arising mutants. 
Individuals are located on the nodes of these graphs, competing with each other to eventually take over the graph via the links. 
Many applications for this framework can be envisioned, 
from the ecology of river systems and cancer initiation in colonic crypts to biotechnological search for optimal mutations.
In all these applications, it is not only important where and when novel variants arise and how likely it is that they ultimately take over, but also how long this process takes. 
More concretely, how is the probability to take over the population related to the associated time? 
We study this problem for all possible undirected and unweighted graphs up to a certain size. 
To move beyond the graph size where an exhaustive search is possible, we devise a genetic algorithm to find graphs with either high or low fixation probability and 
either short or long fixation time and study their structure in detail searching for common themes. 
Our work unravels structural properties that maximize or minimize fixation probability and time, which allows us to contribute to a first map of the universe of evolutionary graphs. 

\newpage

\section{Introduction} 
How does population structure affect evolutionary dynamics?
This question is at the center of evolutionary graph theory, 
since its introduction by Lieberman et al.~\cite{lieberman:Nature:2005}
a flourishing research topic \cite{nowak:book:2006,sood:PRE:2008,broom:PRSA:2008,broom:Evol:2012,monk:PRSA:2014,adlam:SciRep:2014,pattni:PRSA:2015,jamieson-lane:JTB:2015,adlam:PRSA:2015,kaveh:JRSOS:2015,hindersin:PlosCB:2015,pavlogiannis:COMMSB:2018}.
As usual, we will focus on two types only, a mutant and a resident. 
Mutants are assigned a relative fitness $r$ compared to the resident's fitness $1$. In every time step, an individual is chosen for reproduction and its offspring replaces another random individual chosen for death.

More formally, we use the Moran process as a model for invasion of a mutant type into a resident population \cite{moran:PCPS:1958}.
The original Moran process that does not include population structure serves as a reference case reflecting the well-mixed population.
In the case with population structure,
the individuals of the population are located on the nodes of the graph  \cite{lieberman:Nature:2005}.
Different from the standard Moran process, replacement is not taking place among the whole population, but only among the neighbors of the reproducing individual \cite{lieberman:Nature:2005,nowak:book:2006,broom:PRSA:2008,broom:book:2013,maciejewski:JTB:2014}.
Different updating mechanisms are of interest, as they can strongly influence the results \cite{hindersin:PlosCB:2015}:
In Birth-death updating (Bd), a random individual is chosen for reproduction with probability proportional to its fitness to produce an identical offspring, which then replaces a random neighbor.
In death-Birth updating (dB), a random individual dies and this vacant site is immediately filled by the offspring of one of its neighbors with a probability proportional to their fitness.
There are many more possible updating mechanisms, e.g.~bD and Db, where fitness is attached to the death step, and it has been shown repeatedly that the details of this implementation can have a major influence on the dynamics of the process \cite{antal:PRL:2006,altrock:PRE:2009,zukewich:PlosOne:2013,maciejewski:JTB:2014,debarre:NatComm:2014,kaveh:JRSOS:2015,hindersin:PlosCB:2015}. 
Here, we are focussing on Bd updating as the most popular version.

The complete graph, corresponding to the well-mixed population, serves as a reference case to which other graphs can be compared with respect to their fixation probability and time.
Amplifiers and suppressors of selection are graphs that differ from the complete graph in their fixation probability in a particular way \cite{lieberman:Nature:2005,broom:PRSA:2008}:
An amplifier of selection in the most strict sense is a graph, where an advantageous mutant ($r>1$) has a higher fixation probability and a disadvantageous mutant ($r<1$) has a lower fixation probability than on the complete graph. A suppressor of selection is the reverse.

For weighted graphs, it has been shown that graphs that do not differ too much from the complete graph, also have a similar fixation probability \cite{adlam:SciRep:2014}.
More recently, it has been shown that some directed graphs can be strong amplifiers of selection \cite{pavlogiannis:COMMSB:2018}. 
However, strong amplification typically comes at a cost: 
Many amplifiers of selection have been found to slow down the fixation process by increasing the time to fixation \cite{broom:PRSA:2010,frean:PRSB:2013,hindersin:JRSI:2014,tkladec:arxiv:2018}.
For example, Frean et al.~showed that the average conditional fixation time is much higher on the star than on the complete graph \cite{frean:PRSB:2013}.
Even on graphs such as the cycle, which have the same fixation probability as the complete graph, fixation time is increased.

In \cite{hindersin:PlosCB:2015}, it was shown numerically that  most undirected random graphs (up to size $N=14$) are amplifiers of selection for Bd updating and uniform placement of the mutant.
For each $N$, a large number of Erd\H{o}s-R\'enyi graphs~\cite{erdos:PMHAS:1960} were generated and their fixation probability was calculated.
Using improved numerical methods \cite{hindersin:BioSys:2016}, it is possible to study {\em all} graphs up to a certain size instead of looking at the subset of random graphs and to classify them in terms of fixation probability and time. 
However, the space of all graphs increases rapidly, therefore different approaches are necessary to search through it.
Starting from size $N=11$ (more than 1 billion connected undirected graphs \cite{sloane:URL:2009}), we employ a genetic algorithm to optimize for certain properties. This approach allows us to infer if the graph features suggested to optimise e.g.\ fixation probability are still optimal in much larger graphs. 
In particular, we focus on a search of those graphs with the highest and lowest fixation probability and time for a given $N$ and $r$.
We cannot guarantee that the genetic algorithm actually finds the global optimum.
But for sizes up to $N=10$, we confirm that the genetic algorithm finds exactly the same optima that we find by systematically scanning through all graphs.
This worked for all the directions: minimizing and maximizing fixation probability and conditional fixation time.
This serves as a proof-of-concept for the application of the genetic algorithm and suggests that it works reasonably well for larger graph sizes as well.

\section{Results}

\subsection{Correlation of fixation probability and time}

\begin{figure}[!h]
	\centering
	\includegraphics[scale=0.13]{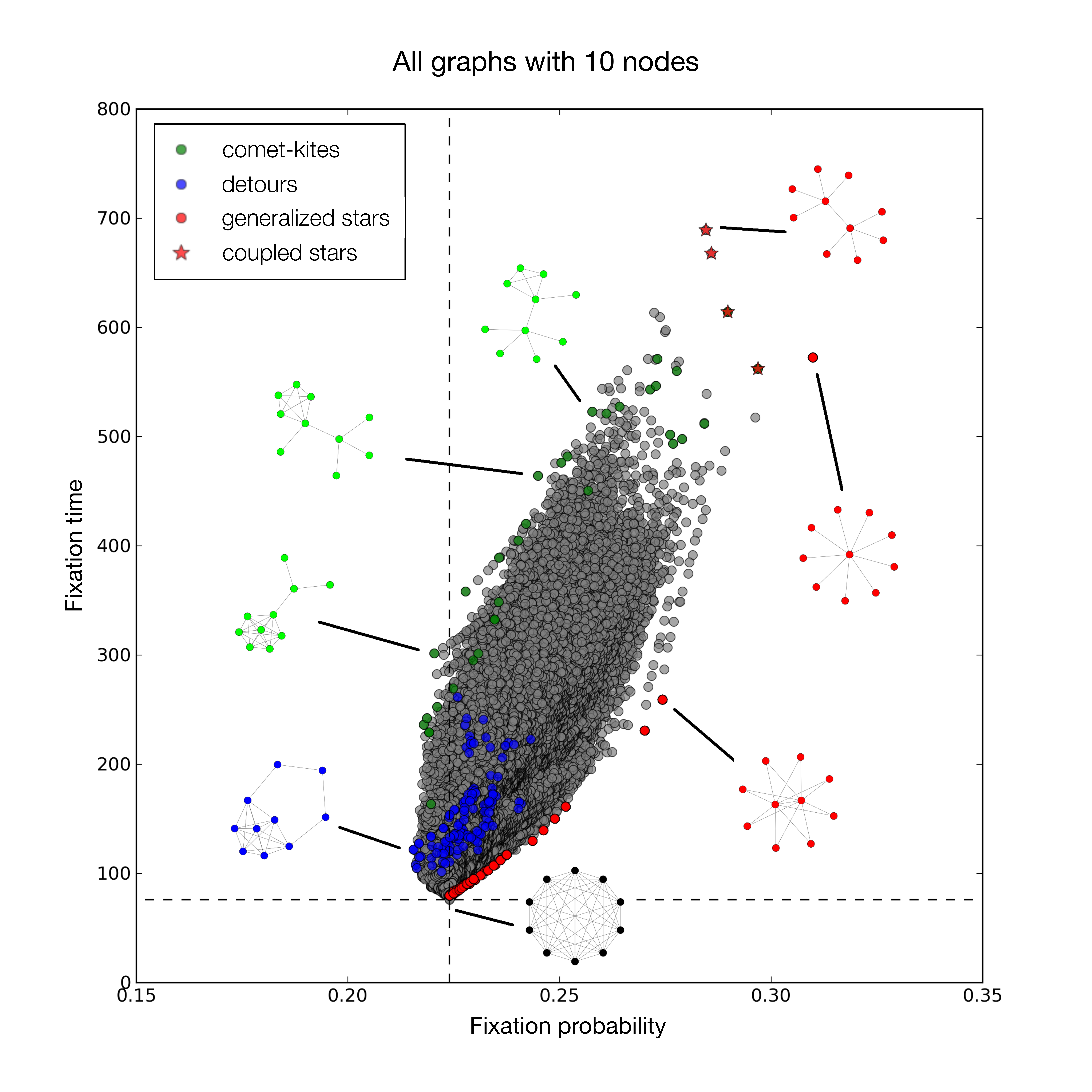}
	\caption{{\bf All graphs of size 10.}
	Overview showing the fixation probability and mean conditional fixation time of all graphs with $10$ nodes as grey dots for $r=1.25$. Certain special types of graphs are portrayed showing their structure. The dashed lines indicate the fixation probability and time for the complete graph.
We highlight three categories of graphs: 
The red graphs are generalized star graphs that provide high fixation probability to the mutants. A subcategory of these generalized stars are coupled stars with the highest fixation times.
The blue graphs are ``detour" graphs that tend to have low fixation probability and time.
The green ``comet-kite" graphs represent part of the left border of the set of all graphs. They minimize fixation probability while spanning a large range of fixation time values.
}
	\label{fig:overview}
\end{figure}

Modifying the graph structure to increase the probability that a mutation takes over often comes at a cost: The modification can at the same time increase the time it takes for the mutant to take over. 
To explore this issue, we visualize all graphs of size $N=10$ in the plane of fixation probability and mean conditional fixation time for particular fitness values.
Fig.~\ref{fig:overview} shows that probability and time are highly correlated.
For example, in biotechnology one could be interested in designing a system that either amplifies or suppresses advantageous mutants. This could be achieved by graphs that have a high fixation probability and a low fixation time or vice versa. In this context, the correlation between probability and time can be seen as a kind of tradeoff.

In addition, Fig.~\ref{fig:overview} highlights special categories of graphs with extremal properties. 
We describe in \nameref{sec:methods} how these categories are generated.
Some particular graphs are of special interest: 
For example, compare the coupled star, which arises from coupling two smaller stars, and the generalized star with two central nodes. 
From a structural point of view, they only differ in that the leaves are connected to both centers instead of to only one, respectively. 
In terms of fixation probability, they are almost identical, but the coupled star has a much higher fixation time.
As another example, the generalized star with two central nodes can also be compared to the comet-kite graphs with relatively low fixation time. 
They have an almost identical fixation time, but are very different in fixation probability. 
Structurally, they are also quite distinct: The comet-kite graphs have a large region of fully connected nodes, and only a few tail nodes, 
while the generalized star has a very small central region. 
The star graphs, the comet-kites and the detour graphs cover the boundaries in terms of fixation probabilities and time, but one has to keep in mind that 
Fig.~\ref{fig:overview} shows only one particular value of $r$. 
Fig.~\ref{fig:overview6and8} displays all graphs of the smaller sizes $6$ and $8$ in the probability-time-plane for different fitness values $r$.
In the supplement, we also provide a .gif file which shows how the graphs move in
the plane with increasing $r$. 
The graphs can be clustered in similar categories as for size $10$, with changes for the extreme graphs.
For disadvantageous mutants with fitness value below $1$, here $r=0.5$, the categories are roughly mirrored on the vertical line that represents the complete graph.

\begin{figure}[!h]
	\centering
	\includegraphics[width=\textwidth]{readyFigures/Overview6And8}
	\caption{
{\bf Changing graph properties with changing fitness.}	
	Overview showing the fixation probability and mean conditional fixation time of all graphs with $6$ and $8$ nodes as grey dots for different fitness values. The dashed lines indicate the fixation probability and time for the complete graph. Special graph categories are again shown in colors (cf.\ Fig.~\ref{fig:overview}.)
Light grey	shows those graphs that have a least one node with a single link,
dark grey shows all remaining graphs. It turns out that this distinction can
capture the separation of  all graphs into two regions to some extent. 
In the supplement, we show an animation of this figure for changing $r$ as a .gif 
	}
	\label{fig:overview6and8}
\end{figure}

\subsection{The strongest undirected suppressors of selection}

Let us now focus on graphs suppressing selection, i.e.\ reducing the fixation probability compared to the fully connected case. 
With our numerical algorithm, we can systematically search for the graphs with the lowest fixation probability for particular values of $r$. 
Studying all graphs up to size $N=10$,
we find that the same graphs are the strongest suppressors  for different values of $r$. 
 Fig.~\ref{fig:supp}a) shows these strongest suppressors for sizes $5$ through $10$.  Their structure is very distinguished through a ``core" part connected by a ``detour".
They resemble a mixture of a completely connected part with a cycle part.
Interestingly, their mean conditional fixation time lies between the complete graph and the cycle as well.

For larger population sizes, we must resort to a genetic algorithm,
as an exhaustive analysis of all graphs is no longer feasible (see \nameref{sec:methods}). 
In a nutshell, we use this algorithm to select on particular graph properties in a population
of graphs. We then produce offspring with some perturbations (motivated by mutation
and recombination) and proceed until the graph property does not change anymore. 
Using this genetic algorithm to search for minimal fixation probability of the graph sizes $6-10$
leads to the same graphs as by systematically searching through all graphs, 
validating our approach. 

For larger sizes of $11-15$, the similar structures remain the strongest suppressors for $r=1.25$, but some details change. 
For example, at size 15 the connection between the core and the detour becomes smoother by having more intermediate nodes which belong to neither group, see Fig.~\ref{fig:supp}b).
Moreover, the genetic algorithm also reveals that the optimum can change with $r$. 
While the overall characteristic of a dense core with a detour remains important for a low fixation probability, additional detours can appear 
for $r=2.0$, cf.\  Fig.~\ref{fig:supp}b).

\subsection{The strongest undirected amplifiers of selection}

We can employ the same approach and search for the highest fixation probability instead. For small $N$, a systematic search shows that the star has the highest fixation probability. 
For sizes of $11-16$, the star is identified again as the strongest amplifier. 
Of course, stronger amplifiers of selection are already known \cite{lieberman:Nature:2005}, but they are all directed or weighted graphs.
Pavlogiannis et al.\ found a class of undirected graphs that they call comets which can have a higher fixation probability than stars for certain graph sizes of more than 100 nodes \cite{pavlogiannis:SciRep:2017}.
Comet-graphs consist of a clique with a star attached to one of the clique-nodes.

\subsection{Optimizing the fixation time}

Instead of focussing on the probabilities of fixation, we can also 
consider the time to fixation. 
This time is randomly distributed, with
a theoretical lower bound of $N$ time steps from the emergence of the mutant to fixation. 
As usual, we focus on the average time to fixation here. 
In our systematic search, the complete graph always emerged as the graph with fastest average time to fixation. However, the proof that the complete graph represents the global minimum for the average fixation time for any graph size is still an open challenge.

Alternative, we could search for the graphs that maximize the time to fixation. 
These are typically star like structures, which offer many weakly connected nodes where the type of the node remains unchanged for very long time.
The precise structure of this graph depends on both the size of the graph $N$ and the fitness advantage $r$.  
For example, for $N=10$ and $r=1.25$, we find a coupled star as the slowest structure, cf.\ Fig.~\ref{fig:overview}.

\begin{figure}[!h]
	\centering
	\includegraphics[scale=0.06]{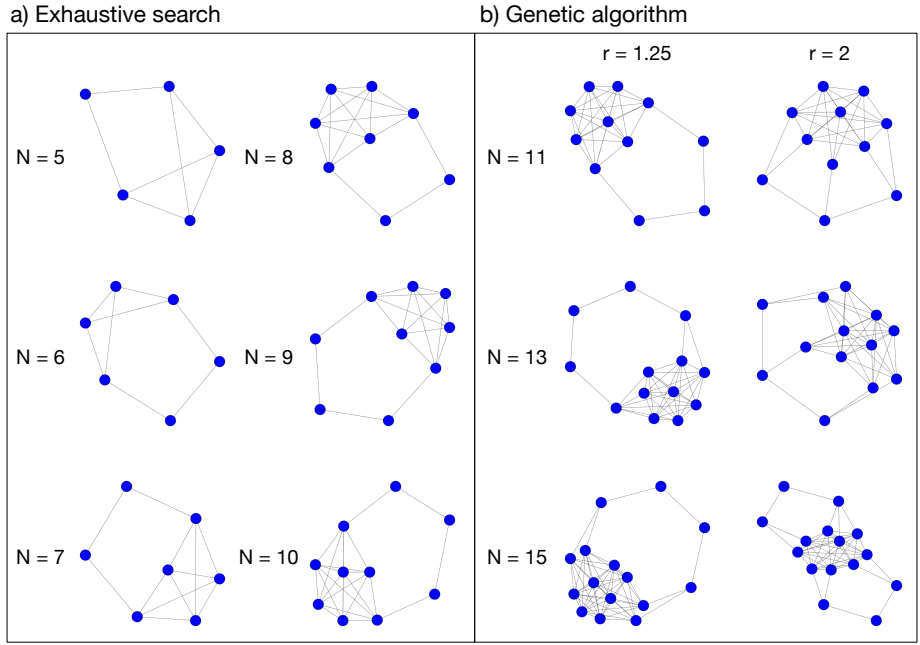}
	\caption{
{\bf Graphs with the lowest fixation probability under Bd updating. }
a) For small $N$, we can exhaustively search through all graphs of the respective size and find that for $r=0.5$ (where the fixation probability is higher than for the complete graph), $r=1.25, 1.5$, and $1.75$,  the strongest suppressors are always identical. Interestingly, for weaker selection ($r\approx 1$) some kite-like graphs are the strongest suppressors.
b) For larger $N$, such a systematic search is no longer feasible. 
Thus, we employ a genetic algorithm to minimize the fixation probability. 
The strongest suppressors found in this way for $r=1.25$ and $r=2$ shown here are structurally very similar to the smaller structures depicted in a).}
	\label{fig:supp}
\end{figure}

\subsection{Changing graph properties with $r$}

For our numerical approach, we must naturally focus on particular values of $r$ and particular graph sizes. 
Classifying a graph numerically as an amplifier or suppressor of selection in this way can never be a proof, as only a finite number of $r$ values can be tested. In a previous numerical study, we used five values of $r$  \cite{hindersin:PlosCB:2015} to classify graphs, as for those graphs 
a higher number of values for $r$ did not change the results in any way.
A recent paper confirms the large proportion of amplifiers with exact symbolical calculations for size $6$ and $7$ and numerics with small step size in $r$ for sizes up to $10$ \cite{cuesta:bioRxiv:2018}.
The vast majority of graphs falls into the three classes of (i) regular graphs (which have the same fixation probability as the complete graph~\cite{lieberman:Nature:2005}), (ii) amplifiers and (ii) suppressors of selection. But some interesting exceptions exist, e.g.\ graphs that reduce the fixation probability for both advantageous and disadvantageous mutants \cite{hindersin:BD:2016}, and graphs that exhibit transitions from being a suppressor to being an amplifier for increasing $r$ \cite{cuesta:bioRxiv:2018}.
In our context, we cannot be entirely sure that the graph properties
stay the same when the fitness value $r$ is changed: A graph could reduce the fixation probability for certain values of $r$, but increase it for others. 
It turns out that there are several graphs that are neither a true amplifier nor a true suppressor of selection~\cite{cuesta:bioRxiv:2018}.
For example, it has been shown that the cotton-candy graph of size $N=10$ (a kite with tail length 1) is a piecewise suppressor~\cite{choi:PA:2018}.
Cuesta et al.\ found a class of graphs called $l$-graphs, which are suppressors of selection for size up to $N=10$ and for all $r$ \cite{cuesta:PLoSOne:2017}. 
The associated mathematical proof becomes challenging for larger $N$, but it is complemented by a numerical exploration that indicates that the same result holds for $N$ up to $24$.
These $l$-graphs are structurally similar to the graphs in Fig.~\ref{fig:supp}, their ``detour" is simply a link connecting two nodes that are each connected to half of the ``core".

\begin{figure}[!h]
	\centering
	\includegraphics[scale=0.17]{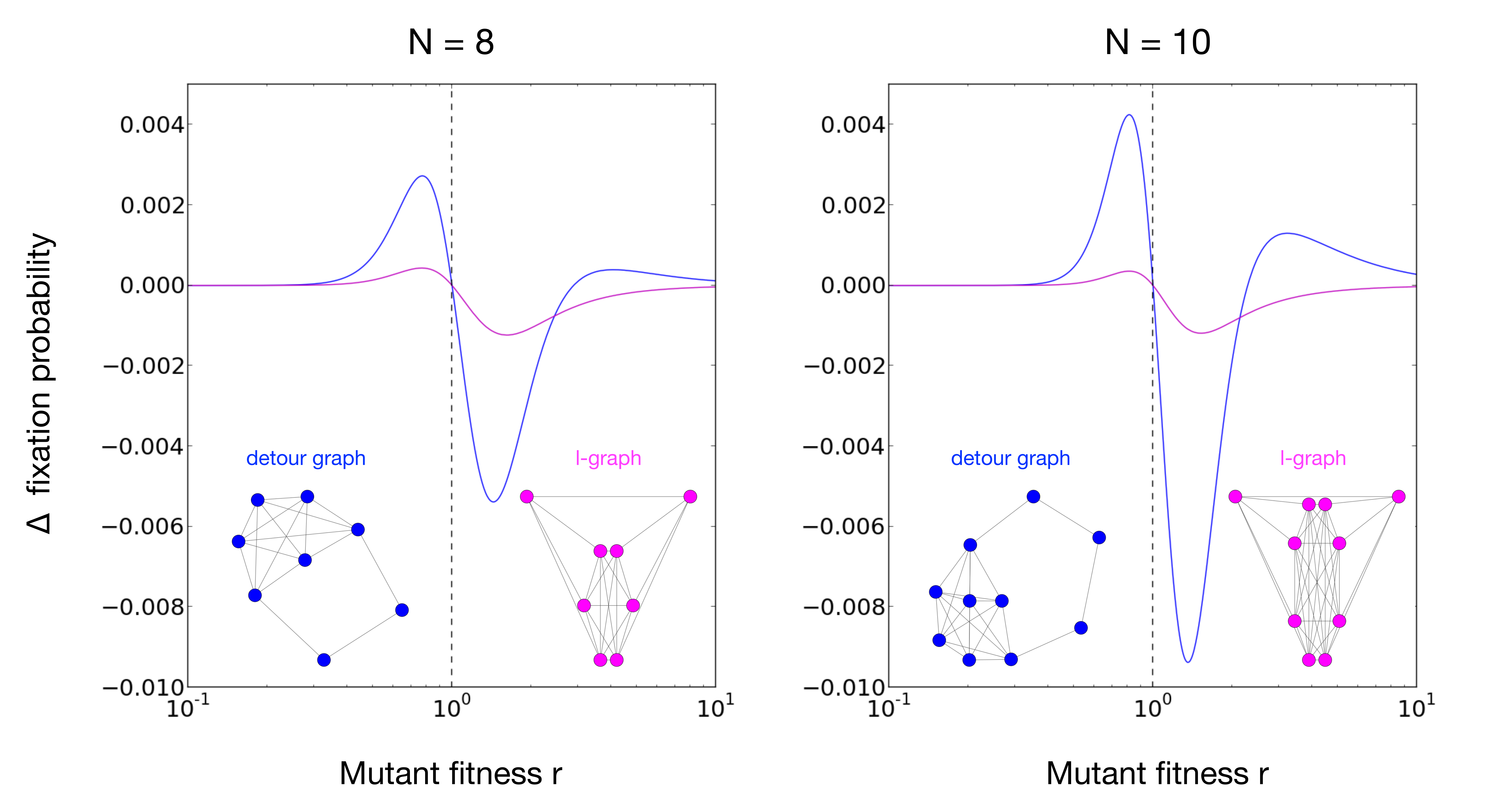}
	\caption{{\bf Switching between amplification and suppression with changing fitness.}
	Up to size $N=10$, $l$-graphs are suppressors of selection for any fitness value $r$. 
	In contrast, detour graphs switch from suppression to amplification at a critical value of $r$.
The figure shows the difference between the fixation probability from the given graphs to the complete graphs of the same size. 
The blue line represents the detour graph and the magenta line shows the result for the $l$-graph.
	}
	\label{fig:r_vs_phi}
\end{figure}

Fig.~\ref{fig:r_vs_phi} shows that the detour graphs are actually piecewise suppressors of selection.
They are very strong ``suppressors" up to a certain $r$ and then they transition into being an ``amplifier".
We put the quotation marks, because these graphs fulfill neither the original definition of a suppressor, nor the original definition of an amplifier.
The $l$-graphs on the other hand are true suppressors of selection, their fixation probability is below the one of the complete graph for all $r>1$ and above the complete graph for all $r<1$ \cite{cuesta:PLoSOne:2017}.

\FloatBarrier
\newpage

\section{Discussion}

Recently, there have been proposals to optimize graph structures to amplify selection in favour of fit mutants.  
Pavlogiannis et al.\ have shown that arbitrarily strong amplifiers of selection can be constructed from almost any graph, as long as self loops and weighted links are allowed \cite{pavlogiannis:COMMSB:2018}. 
For undirected and unweighted graphs, there is much less freedom to construct strong amplifiers. 
However, the simplicity of undirected unweighted graphs allows to ask which structural properties make a graph a strong amplifier of selection. For random graphs, it has been shown that the heterogeneity of a graph, given by the variance between the average speed at which nodes are replaced, is strongly correlated with the fixation probability \cite{tan:SciRep:2014}. In the Appendix, we show that this correlation also holds for all graphs of smaller size:
To construct strong amplifiers of selection, it makes sense to start from graphs that are highly heterogenous.

However, as we can see in Fig.~\ref{fig:overview}, there is a conspicuous correlation between the fixation probability and the average fixation time. 
A naive optimization of fixation probability on the space of all possible undirected and unweighted graphs without self loops always yields the star (for the sizes and fitness values we studied), which also happens to be one of the slowest fixating graphs. 
We considered both properties at once, to discern what properties lead to a graph fixating slowly or fast, or being a strong amplifier or suppressor. 
In particular, we found some categories of graphs that optimize both the fixation probability and the fixation time at the same time.
Tkladec et al. call the graphs where the fixation probability cannot be increased without increasing the fixation time ``Pareto-optimal'' \cite{tkladec:arxiv:2018}.

When maximizing fixation probability and minimizing fixation time, we found the category of the generalized star \cite{houchmandzadeh:BioSys:2013}, to which the star belongs as well. 
However, all the other graphs in this category have a much lower fixation time than the star, making certain ones possibly better when constructing an environment where you want mutants with higher fitness to fixate both fast and reliably.

On the other hand when minimizing fixation probability and maximizing fixation time, we obtain the ``comet-kites''.
They look like kites with a few extra, comet-like tails. 
When maximizing fixation time alone, we obtain coupled stars instead.

Depending on which property is considered more important, a wide varity of different graphs can be optimal.
Furthermore, we showed how a genetic algorithm can be used to optimize for certain properties. 
Not only was the genetic algorithm capable of finding the graphs for the lower sizes which we already knew to be optimal, it also could find interesting graphs for higher sizes.
This approach can be expanded to look for a combination of fixation probability and time, or for completely different directions.
With the fast computational method we use for computing fixation probability and time, we are limited to sizes roughly up to $N=23$ \cite{hindersin:BioSys:2016}. 
There are many interesting findings for larger graphs, e.g.\ that the star is not the strongest amplifier for all sizes, but that comets can have an even higher fixation probability \cite{pavlogiannis:SciRep:2017}.
Our genetic algorithm can be easily combined with other methods of computation or simulation and to tackle larger graph sizes.

Many open questions remain in the field of evolutionary graph theory. 
Our approach of looking at both the probability and the time of fixation 
reveals interesting graphs at the edges of optimality. 
With such approaches, the construction of a map of all evolutionary
graphs can be envisioned. 

\section{Methods}
\label{sec:methods}

\subsection{Computing fixation probabilities and times}

In order to compute the fixation probabilities and times, we use
the adjacency matrix based approach published in \cite{hindersin:BioSys:2016}. Note that this approach does not exploit and symmetries of the graphs and is thus limited by the population size to $N\approx 23$. 

\subsection{Generating all graphs}

To numerically generate all possible graphs up to a certain size, we use the software geng from the package nauty \cite{mckay:JSC:2014}. 
In particular, we can generate all connected, undirected graphs of a certain size with ``geng size -c'', which can can be converted to a more easily readable format with a program from the same package, showg. \\
This can then be used for a shell script or a python script.

\subsection{Drawing the graphs}

We use the Kamada-Kawai force-directed algorithm to plot graphs \cite{kamada:IPL:1989}. 
This algorithm draws graphs in a manner which clearly presents structure.

\subsection{Generating special graph categories}
\label{subsec:categories}
Here, we describe how we generate the categories for Figs.~\ref{fig:overview} and \ref{fig:overview6and8}.

\paragraph{Comet-kites} \mbox{}\\
Input parameters: \\
$N$: Number of nodes in the graph (integer)\\
$c$: Number of central, fully connected nodes (integer)\\
$t$: Number of comet-like tails connected to one of the central nodes. (integer)\\

\noindent
Algorithm:
\begin{enumerate}
\item Fully connect the first $c$ nodes to generate a clique (complete sub graph) of size $c$.
\item Attach a number of $t$ nodes to node number $c$ in a star-like fashion.
\item Iteratively add one link respectively from every remaining $N - c-t$ nodes randomly somewhere to the tails. 
\end{enumerate}

\begin{figure}[!h]
	\centering
	\includegraphics[width=\textwidth]{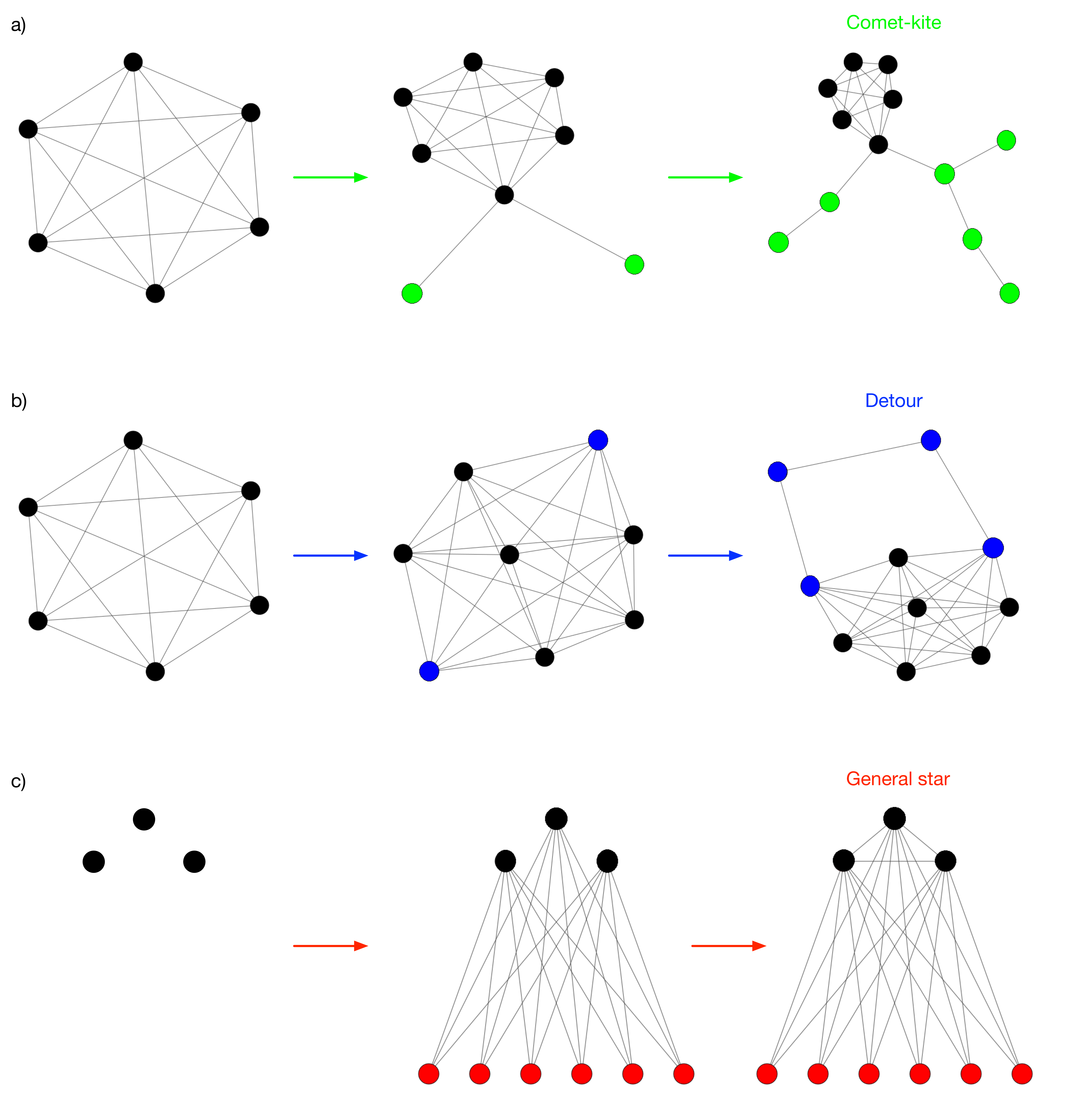}
	\caption{
	{\bf Examples for the graph generator algorithms.}
	a) The three steps of the cometkite-algorithm. First the fully connected region (in black) is generated, then the ``roots'' of the ``tails'' (in green), then the random additions to the tails (also in green).
	b) The three steps of the detour-algorithm. First the fully connected region (in black) is generated, then the outer region (in blue), then the detour (also in blue).
	c) The two steps of the general star-algorithm. First the bipartite graph (first partition shown in black) is generated, then connections between the nodes in the first partition are added.		
	}
	\label{fig:generator_algorithms}
\end{figure}

This process shown in Fig.~\ref{fig:generator_algorithms}a)
generates a single ``comet-kite" graph. 
For the figures, we varied $c=1,2,...,N-1$ and $t=1,2$ for a given $N$. 
Higher numbers of tails resulting in more comet-like graphs were less extreme in terms of fixation probability and time.
Additionally, the standard kites are added, which are the graphs with a fully connected region and just a single line connecting to one of the nodes. 
This is done because it is unlikely to generate them by chance.

\FloatBarrier

\newpage
\paragraph*{Detours} \mbox{}\\
Input parameters: \\
$N$: Number of nodes in the graph (integer)\\
$c$: Number of central, fully connected nodes (integer)\\

\noindent
Algorithm:
\begin{enumerate}
	\item Fully connect all nodes up to node $c$ to generate a clique (complete sub graph) of size $c$, which forms the inner region of the graph.
	\item Generate a random integer $o \in \{2,\textrm{min}(c,N-c-1) \}$, which is the number of outer region nodes. 
	This range of values is chosen to avoid that the outer region is larger than the inner region and to ensure that there are enough remaining nodes for the detour.
	\item Generate the detour. This means adding links from the remaining $N-c-o$ nodes randomly somewhere to the outer region or other nodes in the detour until all nodes in the detour have at least 2 connections.
	\item Check whether the outer region nodes are connected with the detour. If not, repeat.
\end{enumerate}
This generates a single ``detour" graph, the algorithm is depicted in Fig.\ \ref{fig:generator_algorithms}b). We vary $c$ from $c=2,3,...,N-4\}$.
In addition, we add the ``standard'' detour graphs, which are those graphs where $o = 2$ and where the detour is connected in a ring-like fashion from one of the outer region nodes to the other.

\paragraph*{General stars} \mbox{}\\
Input parameters: \\
$N$: Number of nodes in the graph (integer)\\
$a$: Number of nodes in the first partition (integer)\\
$p$: Likelihood of the nodes in the first partition to be connected to each other (real number). If $p=0$, we have a standard bipartite graph. If $p=1$, the first partition is fully connected.\\
\begin{enumerate}
	\item Generate a bipartite graph with $a$ nodes in one partition and $N-a$ nodes in the other, where all nodes in one partition are connected to all nodes in the other partition. 
	\item Randomly connect the nodes within the first partition with probability $p$.
\end{enumerate}
This generates one ``generalized star" graph. We vary $a \in \{1,\ldots, \lfloor N/2 \rfloor \}$ and $p \in \{ 0.01, 0.02, \ldots,1 \}$ for a given $N$.\\
This process is shown in Fig.~\ref{fig:generator_algorithms}c).

\paragraph*{Coupled stars} \mbox{}\\
Input parameters: \\
$N$: Number of nodes in the graph (integer)\\
$a$: Number of nodes connected to the first center of the coupled star. (Integer)\\
\begin{enumerate}
	\item Connect the first two nodes to each other. These are the central nodes.
	\item Connect $a$ nodes to the first node and $N-a-2$ nodes to the second node.
	
\end{enumerate}
We vary $a \in \{0,1,..., \lfloor (N-2)/2 \rfloor \}$ for a given $N$.

\subsection{The genetic algorithm}
\begin{figure}[!h]
	\centering
	\includegraphics[scale=0.12]{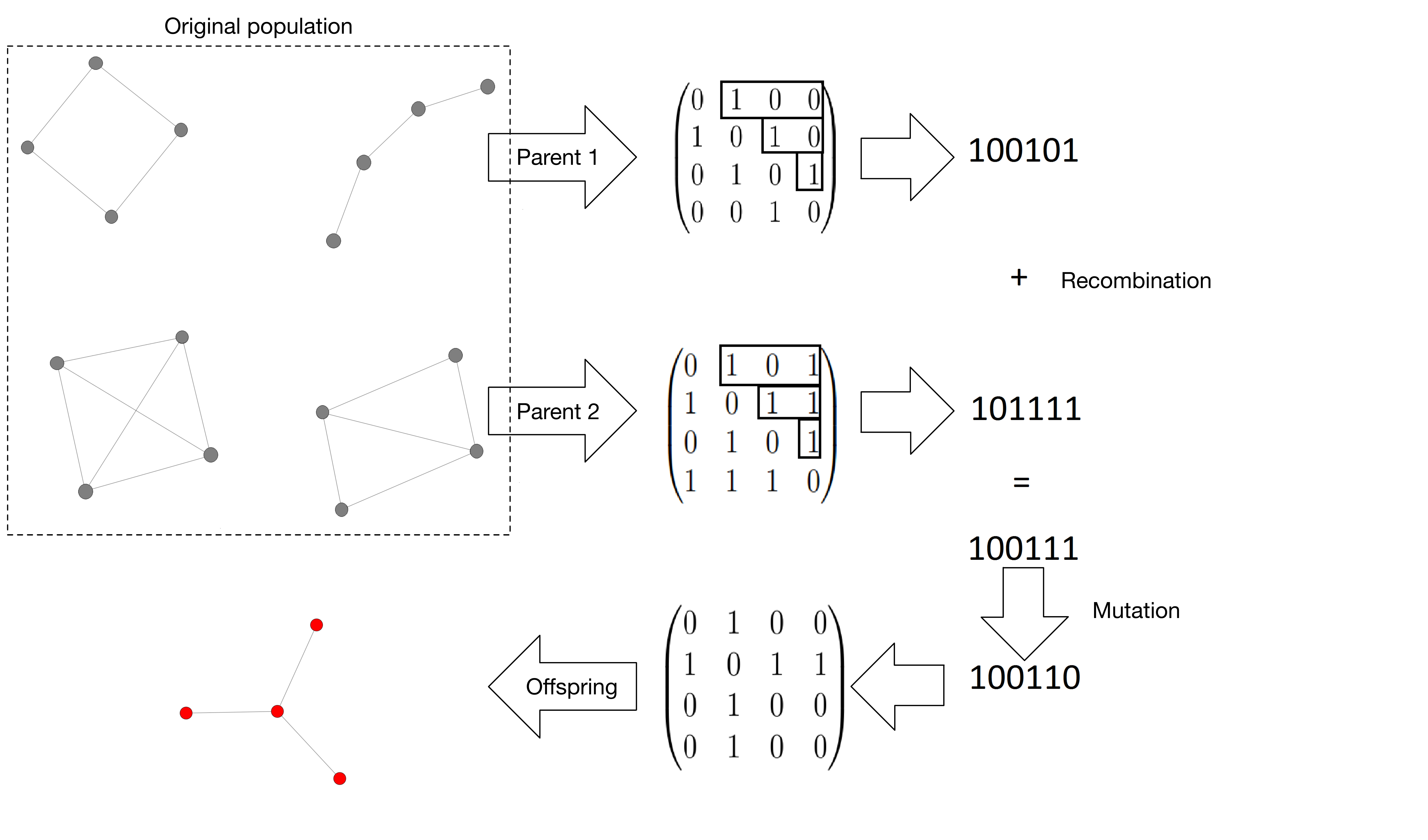}
	\caption{
	{\bf Illustration of the genetic algorithm.}
	A single step of the genetic algorithm for the case of optimizing the fixation probability. Out of a population of graphs, the algorithm chooses two with the highest fixation probability and converts them to the gene code. The gene codes are recombined and mutated and reverted back into an adjacency matrix.
	}
	\label{fig:genalgo}
\end{figure}

For small $N$, when looking for extreme graphs, we simply generated all graphs for a given size and then chose the ones with interesting properties. 
This, however, is not possible anymore for increasing sizes, since not only does the wall time per graph increase exponentially, but so does the number of graphs.

Using the Erd\H{o}s-R\'enyi algorithm to generate a huge number of graphs and search through them also does not seem promising, since, as seen in Fig.~\ref{fig:overview}, the extreme graphs have very special degree distributions, while the Erd\H{o}s-R\'enyi algorithm strongly tends towards binomial degree distributions \cite{albert:RMP:2002}. 
It is therefore very likely to miss them entirely.

We thus propose a heuristic approach based on a genetic algorithm. In Fig.~\ref{fig:genalgo}, a short overview of the procedure is shown. It can be summarized as following:

\paragraph*{Parameters:} \mbox{} \\
$m$: Number of random graphs, here chosen as 120 \\
$k$: Number of parents per generation, here chosen as 20 \\
$b$: Number of mutations per individual per timestep, here chosen as 1\\
$n$: Number of iterations, here chosen as 5000\\

\begin{enumerate}
	\item Generate $m$ random Erd\H{o}s-R\'enyi graphs.
	\item Calculate the property that we want to optimize for all of the graphs.
	\item Choose the best $k$ graphs, based on the property, as parents for the following generation.
	\item Generate m new graphs, each with two parents, by recombination. Every link can be inherited from either parent with a probability of 50 $ \% $. 
	\item Mutate these graphs with an average of $b$ mutations per individual, based on a binomial distribution. A mutation means here that a random link is either eliminated or created.
	\item Repeat steps 2.\ to 5. many times.
\end{enumerate}

With this approach we only need to look at some relevant graphs for the property we aim to optimise, as opposed to looking at a huge number of graphs. 
We have tested that this genetic algorithm recovers the optima for small graph sizes found by systematically scanning all graphs.

\section*{Acknowledgements}
We thank Christoph Hauert, Alvaro Lozano Rojo, Yuriy Pichugin and Silvia de Monte for fruitful discussions.


\section*{Appendix}
\label{appendix}

\subsection*{Heat heterogeneity}

Given a graph $G$ with $N$ nodes, let the degree of each node be denoted as $d_1, d_2, \cdots, d_N$.
The temperature of node $i$ is then defined as $T_i = \sum_{k\in N(i)}{\frac{1}{d_k}}$, where $N(i)$ is the set of neighbors of node $i$ \cite{lieberman:Nature:2005,tan:SciRep:2014}.
Tan et al.\ define heat heterogeneity as 
$H_t(G) = \frac{1}{N}\sum\limits_{i=1}^N{(T_i - \bar{T} )^2}$,
where $T_i$ is the degree of node $i\in \{1,\cdots,N\}$ and $\bar{T}$ is the average temperature of graph $G$ \cite{tan:SciRep:2014}.

Tan et al.\ found a correlation between heat heterogeneity and fixation probability for Erd\H{o}s-R\'enyi (ER) graphs \cite{tan:SciRep:2014}.
For graph sizes $4$--$9$ we can study all graphs.
Table \ref{tab:correlationFULL} shows the Pearson correlation coefficient between the heat heterogeneity and the fixation probability for all graphs of sizes $4$--$9$ and fitness values $r=1.25$ and $r=1.75$.
The Pearson correlation coefficient can attain values from -1 to +1, -1 reflecting a strong negative correlation and +1 a strong positive correlation.
The correlations in Table \ref{tab:correlationFULL} are very high.

\begin{table}[h]
	\centering
	\begin{tabular}{l| l| l}
		
		$\mathbf{N}$ & $\mathbf{r = 1.25}$ & $\mathbf{r = 1.75}$\\
		& & \\
		4 & 0.9573 & 0.9671 \\
		5 & 0.9261 & 0.9425 \\
		6 & 0.8579 & 0.8865 \\
		7 & 0.8324 & 0.8646 \\
		8 & 0.8433 & 0.8746 \\
		9 & 0.8690 & 0.8922 \\	
		
	\end{tabular}
	\caption{Pearson correlation coefficient between the heat heterogeneity and the fixation probability for graph sizes $4$--$9$. All graphs are included for the computation of these correlations.}
	\label{tab:correlationFULL}
\end{table}

For larger graph sizes we cannot study all graphs.
Therefore we use the following procedure to generate graphs:
Set a value $p$ for the ER algorithm to be the probability of link connection.
\begin{enumerate}
\item Generate an ER graph $G$ with the given $p$.
\item Extract the degree distribution of $G$ and generate $100$ graphs with this degree distribution.
\item Check for uniqueness of these graphs and remove duplicates.
\item If less than $10$ unique graphs are left in the set, go back to step 1.
\item Compute the fixation probability and the heat heterogeneity for these graphs.
\item Compute the Pearson correlation coefficient of those. 
\end{enumerate}

We repeate this procedure $10$ times and compute the average correlation of these $10$ trials.
The results are displayed in Table \ref{tab:correlation}.

\begin{table}[h]
	\centering
	\begin{tabular}{l| l l| l l}
		
		$\mathbf{N}$ & $\mathbf{r = 1.25}$ & & $\mathbf{r = 1.75}$ & \\
		 & $\mathbf{p=0.5}$ & $\mathbf{p=0.8}$ &$\mathbf{p=0.5}$ & $\mathbf{p=0.8}$ \\
		& & & & \\
		10 & 0.8743 & 0.8399 & 0.9349 & 0.9688 \\
		11 & 0.8769 & 0.8711 & 0.9677 & 0.9865 \\
		12 & 0.9036 & 0.8921 & 0.9680 & 0.9842 \\
		13 & 0.8873 & 0.9211 & 0.9774 & 0.9832 \\
		14 & 0.9218 & 0.9007 & 0.9746 & 0.9855 \\
		15 & 0.9202 & 0.9135 & 0.9732 & 0.9828 \\
		16 & 0.9049 & 0.9083 & 0.9740 & 0.9890 \\
		
	\end{tabular}
	\caption{Correlation between the heat heterogeneity and the fixation probability for graph sizes $11$--$16$. These correlations are computed from 10 trial runs with 100 graphs each, with an identical degree distribution as one randomly generated ER graph of the given p. Details of this procedure are given in the text.}
	\label{tab:correlation}
\end{table}

\begin{figure}[h]
	\centering
	\includegraphics[width=\textwidth]{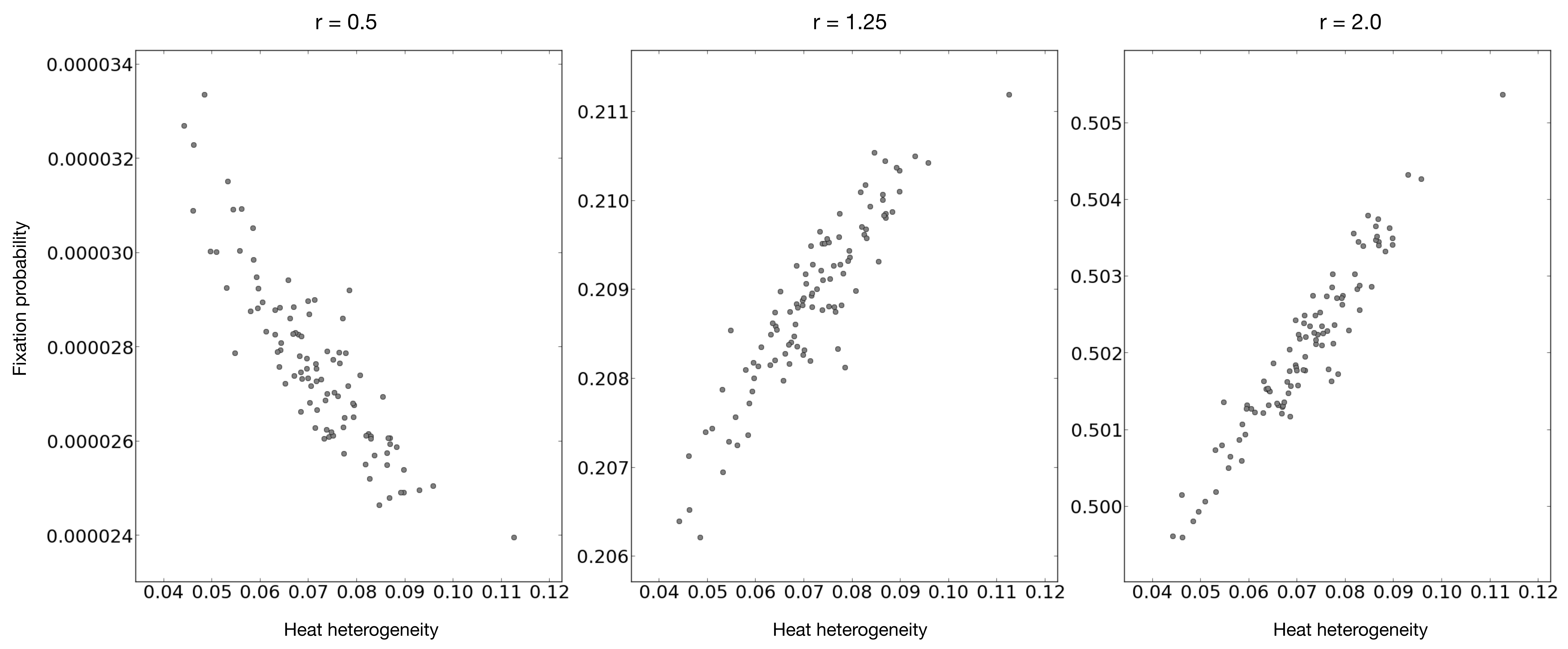}
	\caption{
	{\bf The heat heterogeneity is a good predictor for the fixation probability.}
	The plots show the 
	correlation between heat heterogeneity and fixation probability for size $N=15$ and various values of $r$.
	}
	\label{fig:heatvsfixprob}
\end{figure}

Fig.~\ref{fig:heatvsfixprob} visualizes the correlation of heat heterogeneity and fixation probability for $N=12$ and $r=1.25$.
These results shows that the degree distribution combined with the heat heterogeneity can predict the fixation probability very well.

\end{document}